\documentclass[a4paper,useAMS,usenatbib]{mn2e}
\usepackage{graphicx}
\usepackage{subfig}
\usepackage{ulem}
\title{The JCMT Gould Belt Survey: low-mass proto-planetary discs from a SCUBA-2 census of NGC~1333}
\author[P. Dodds et al.]{P. Dodds$^1$, J.S. Greaves$^1$, A. Scholz$^1$, J. Hatchell$^2$, 
W.S. Holland$^3$ and the 
\newauthor{JCMT Gould Belt Survey Team}\\
$^1$SUPA, School of Physics \& Astronomy, University of St Andrews, North Haugh, St Andrews, Fife KY16 9SS, U.K.\\
$^2$Physics and Astronomy, University of Exeter, Stocker Road, Exeter EX4 4QL, U.K. \\
$^3$UK Astronomy Technology Centre, Royal Observatory, Blackford Hill, Edinburgh EH9 3HJ, U.K.
}

\begin{document}

\date{Accepted 2014. Received 2014; in original form 2014}


\maketitle

\label{firstpage}

\begin{abstract}

NGC~1333 is a 1-2\,Myr old cluster of stars in the Perseus molecular cloud. We used 
850$\,\mu m$ data from the Gould Belt Survey with SCUBA-2 on the JCMT to measure or place
limits on disc masses for 82 Class II sources in this cluster. Eight disc-candidates were 
detected; one is estimated to have mass of about 9 M$_{\rm Jupiter}$ in dust plus gas, while 
the others host only 2-4 M$_{\rm Jupiter}$ of circumstellar material. None of these discs exceeds
the threshold for the 'Minimum Mass Solar Nebula' (MMSN). This reinforces previous claims
that only a small fraction of Class II sources at an age of 1-2\,Myr has discs exceeding the MMSN 
threshold and thus can form a planetary system like our own. However, other regions with 
similarly low fractions of MMSN discs (IC348, UpSco, $\sigma$\,Ori) are thought to be older
than NGC~1333. Compared with coeval regions, the exceptionally low fraction of massive discs 
in NGC~1333 cannot easily be explained by the effects of UV radiation or stellar encounters. Our
results indicate that additional environmental factors significantly affect disc evolution and the
outcome of planet formation by core accretion.
\end{abstract}

\begin{keywords}
protoplanetary discs -- millimetre observations.
\end{keywords}

\section{Introduction}
\label{intro}

Over 1000 exoplanet detections have now been claimed (e.g. exoplanet.eu), ranging from 
single planets to multi-object systems of varied numbers and masses. \citet{2005ApJ...627L.153D} has 
calculated that, based on the rocky content of the solar system, at least 20 
Jupiter masses of gas plus dust are required in a protoplanetary disc to make the 
solar system planets -- the Minimum Mass Solar Nebula (MMSN). Older estimates of 
the MMSN are in the range of 10 to 100 Jupiter masses \citep{1977Ap&SS..51..153W}.

Current searches are finding numerous low-mass planets, in particular around low-mass 
stars, and close analogues to the Sun's system of planets may be uncommon. For example, 
\citet{2013A&A...552A..78Z} estimate the frequency of giant planets with periods 
smaller than 10~yr around solar-like stars to be 10\%, implying that exo-Jupiters are not the 
norm. Using microlensing results, \citet{2012Natur.481..167C} argue that the majority of 
solar-like stars host a Neptunian and/or super-Earth planet, whereas only around 
one-sixth of stars host a gas giant within 10~AU. These results tend to reduce the 
mass requirements for circumstellar discs at early times that can form planetary 
systems via core accretion. \citet{2010MNRAS.407.1981G} have noted that MMSN-discs are 
uncommon around T Tauri stars, and so less substantial planetary systems than that of 
the Sun could be expected to form. Furthermore \citet{2011MNRAS.412L..88G} suggested that dust 
aggregation could begin as early as the protostellar stage, and that Class 0 
discs of $\ga 20$~M$_{\oplus}$ of dust would allow for super-Earths to be common, 
even if only 10 per cent of the solid material is captured into planetary cores.
 
We consider here whether core growth proceeds steadily with time, or whether it is 
also affected by environment. Star formation regions vary greatly, from spare 
associations such as in Taurus-Auriga, up to dense clusters exemplified by the Orion 
Nebula Cluster. Wide-field surveys at wavelengths where dust emission is optically 
thin are thus advantageous, both for covering entire regions and for systematically 
identifying disc signals within complex clouds. One such programme is the Gould Belt 
Survey \citep[GBS;][]{2007PASP..119..855W}, part of the Legacy Project of the James Clerk 
Maxwell Telescope, which examines star formation within $\sim 500$~pc. The SCUBA-2 
submillimetre camera \citep{2013MNRAS.430.2513H} has produced some of the first large-scale
maps of cold dust regions, and is complementary to Gould Belt surveys in the infrared from {\it Spitzer 
{\rm and} Herschel}. We present here early SCUBA-2 850$\,\mu m$ data for the NGC~1333 region, 
while Buckle et al. (in prep.), Broekhoven-Fiene et al. (in prep.) and Drabek-Maunder et al. 
(in prep.) will discuss, respectively, discs in Taurus, in Auriga, and in the wider 
GBS survey.

\begin{figure*}
\includegraphics[width=15cm,angle=-90]{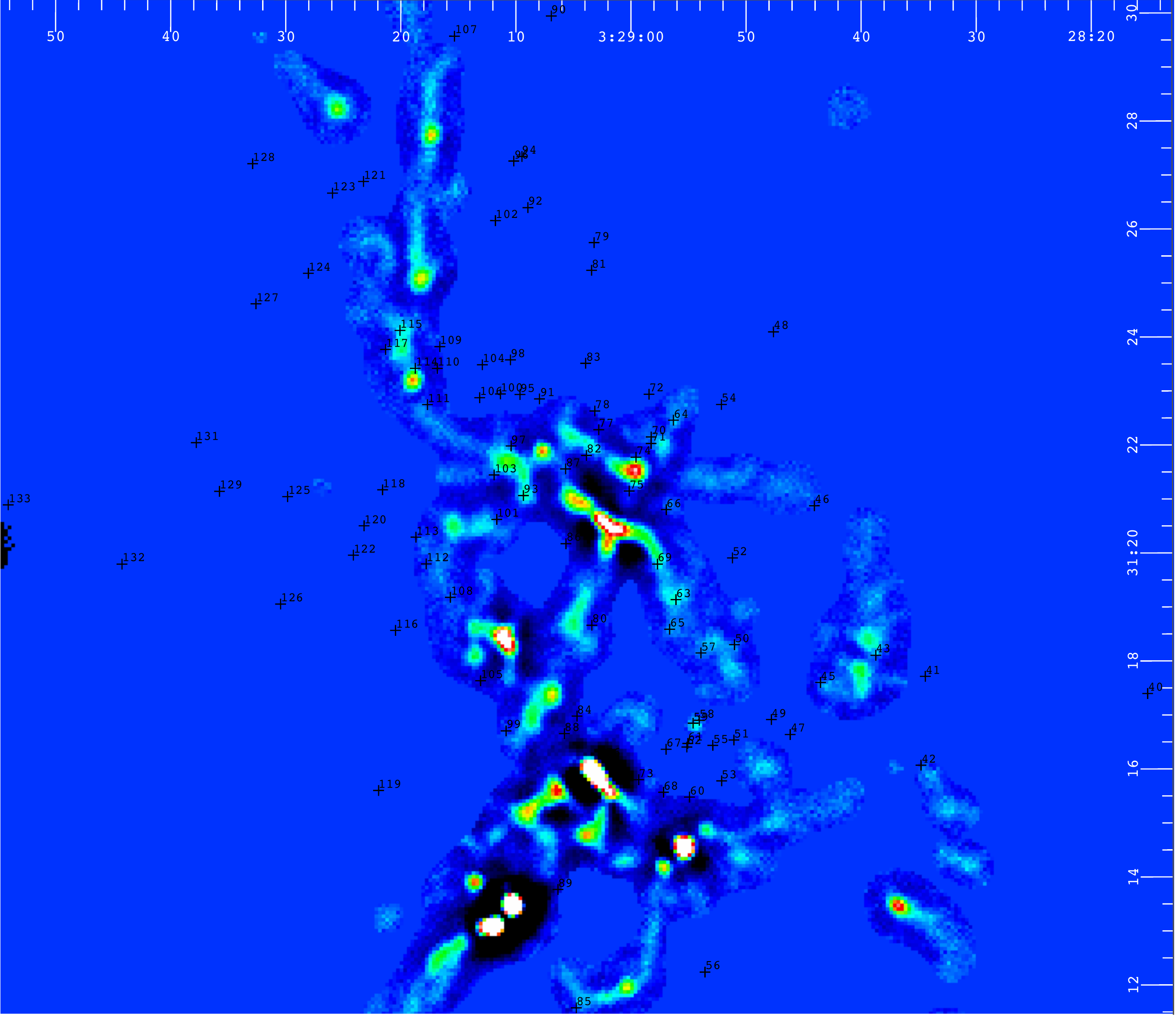}
\caption{Central part of 850$\,\mu m$ map of NGC 1333, after self-subtraction of emission 
of scales $\ga$ 32 arcsec (see text). The false colour scale is linear and ranges from 
--15 to +50 times the per-pixel noise, with dark blue corresponding to zero signal in masked 
areas of the map. Disc candidates are marked with black crosses and annotated with 
their numbers from the Gutermuth et al. (2008) catalogue. One candidate (\#44) is outside
the field of view shown here. \label{map}}
\end{figure*} 

NGC~1333 is a stellar cluster with about 150 young stellar objects, located at approximately 250~pc 
from the Sun, and within the Perseus molecular cloud. Perseus also hosts the IC~348 
region, where 1.3 and 3~mm disc searches have been made by \citet{2011ApJ...736..135L} and 
\citet{2002AJ....124.1593C} respectively. \citet{2013ApJ...775..138S} discuss the star and brown dwarf 
populations of both of these regions. NGC~1333 and IC~348 have estimated ages of 
around 1-2 and 2-4 Myr respectively \citep{2008hsf1.book..308B}. This places them within the 
age bracket of $<1$ 
to 5~Myr, where many young stars host cold discs. The {\it Spitzer} survey of NGC~1333 
by \citet{2008ApJ...674..336G} provides a catalogue for our search for submillimetre dust, 
especially around the Class~II objects that make up two-thirds of this IR-identified 
population. These `classical' T Tauri stars are expected \citep[e.g.][]{1987IAUS..115....1L} to 
have more substantial discs than the remnants around Class~III weak-line sources, without the 
confusion of circumstellar envelopes that dominate in Class I/0 protostellar systems.

\section{Data and analysis}
\label{data}

The observations for NGC1333 were among the first made with SCUBA-2 at the JCMT. The data 
used in this paper consist of 850$\,\mu m$ scan-maps made for the GBS between 21 Feb and 7 
Mar 2010. The same dataset is already discussed in \citet{2013MNRAS.429L..10H}. Data was taken in 
the 'shared risk' campaign; only one sub-array of the final four was available during this campaign. 
The flux conversion factor was $653\pm 49$\,Jy/beam/picoWatt.
The weather conditions were favourable, with
225\,Ghz opacity of $\tau_{225} = 0.05\hbox{--}0.07$. Two overlapping circular regions with
15' diameter were fully scanned, offset by 10' the north and south from 
$\rmn{RA}(2000)=03^{\rmn{h}}~29^{\rmn{m}}$, $\rmn{Dec.}~(2000)=31\degr~18\arcmin$, which is 
close to the center of the cluster \citep{2012ApJ...756...24S}.
The 4' borders around these regions were scanned at lower sensitivity.

The map-mosaic reduction has been optimised to search for dust-disc candidates that 
are point-like within a 14 arcsec beam. As the cloud emission imposes a high dynamic 
range of fluxes, the mosaic was self-subtracted with a version smoothed over a 
Gaussian of 32 arcsec full-width half-maximum. This produces some `bowl' effects 
alongside the bright cloud filaments, but helps in disc identification by flattening 
the background. The basic map processing also includes a masking step, in which 
regions of low signal-to-noise are set to zero, in order to optimise the filtering out 
of artefacts from the scanning process. This masking could potentially result in missing any 
discs that are well separated from the cloud filaments. The smoothing scale of 32
arcsec was chosen to effectively remove background structures; changing this scale
to 24 or 40 arcsec changes the resulting fluxes by less than 10\%.


A 300 square-arcminute area of NGC 1333 was effectively imaged, within which lie 40 of 
the 94 Class II objects found by \citet{2008ApJ...674..336G}. Six of these are in areas 
near filaments and acutely affected by `bowls' and are removed from the sample.
The remaining 34 are typically 30 arcsec away from the 'bowls', which makes reliable aperture 
photometry feasible. Thus, the completeness of the submillimetre disc search is 36\% (34/94). 
For these 34 sources, the map depth is very uniform. The sensitivity at 850$\,\mu m$ 
is 7.8 mJy/beam rms, as measured from the scatter of 20-arcsec pixel boxes in 
the blankest map regions \citep[see also][]{2013MNRAS.429L..10H}. This limit is also typical
for the regions around the Class II sources covered by the map. The images shown have 
4-arcsec pixels, so for aperture photometry, the summed-and-sky-subtracted signals were 
converted from a Jy/beam to Jy/pixel scale. The beam area at 850$\,\mu m$ is 229 arcsec$^2$ 
\citep[effective FWHM of 14.1 arcsec,][]{2013MNRAS.430.2534D}, which is equivalent to 
14 pixels each of $4\times4''$ dimensions. 

As the discs should be unresolved, a small aperture with radius 2.5 pixels was adopted, and 
sky annuli spanned 1.5-2.5 times this radius. Aperture photometry for the 34 Class II sources 
was carried out with the tools of the {\it GAIA} package. The apertures were fixed on the peak
pixel position, without automatic centroiding.
Plausible disc candidates met the criteria of a compact (FWHM $<30$") flux peak located within 
2 pixels (8 arcsec) of the T Tauri position. This roughly half-beam difference between the stellar 
position and centre of the peak pixel allows for centroiding error at low signal-to-noise, plus 
possible pointing drifts of about a half-pixel. The offsets of the peak-flux pixels were found 
to lie in random directions from the target co-ordinates, so pointing systematics are not 
significant.

\begin{table*}
 \begin{center}
  \caption{NGC1333 protoplanetary disc candidates; note that object \#59 is blended with \#58
(Figure \ref{map}) so it is uncertain which source dominates the emission. Object \#59 is located in the pixel of
the peak emission, while \#58 is 7.5 arcsec away. The first five columns list the 
RA-ordered object numbers, RJ co-ordinates, YSO names, K-band extinction and mid-IR spectral 
slope from \citet{2008ApJ...674..336G}. The slope is a power-law index across the 
Spitzer-IRAC 3.6-8$\,\mu m$ bands (not extinction-corrected), with values exceeding --1.8 characterising 
Class II discs. The final two columns list the SCUBA-2 850$\,\mu m$ flux from aperture photometry and 
the inferred disc mass in gas plus dust. See text for discussion of errors. \label{t1}}
  \begin{tabular}{llllrrr}
  \hline
id\#	& RA,Dec			& names				& A$_K$	&$\alpha_{IRAC}$& F850 & M$_\mathrm{{disc}}$\\ 
	&				&				&	&	& (mJy)	& (M$_\mathrm{{Jup}}$)   \\
\hline
 45 	& 03 28 43.56 	+31 17 36.5 	& ASR 127                    	& 0.41 	& -0.24	& 32	& 4		\\
 50 	& 03 28 51.02 	+31 18 18.5 	& SVS 10; ASR 122; LAL 106   	& 0.41 	& -0.62	& 31    & 4		\\
 59 	& 03 28 54.61 	+31 16 51.3 	& SVS 18; ASR 43; LAL 136    	& 0.93 	& -0.70	& 34    & 4		\\
 63 	& 03 28 56.09 	+31 19 08.6 	&                               &      	&  0.47	& 29    & 4		\\
 64 	& 03 28 56.31 	+31 22 28.0 	& LAL 147                    	& 2.19 	& -0.22	& 22    & 3		\\
 65 	& 03 28 56.64 	+31 18 35.7 	& ASR 120; LAL 150           	& 0.80 	& -1.45	& 20    & 3		\\
 93 	& 03 29 09.33 	+31 21 04.2 	& LAL 225                    	& 1.51 	& -0.43	& 74    & 9		\\
111 	& 03 29 17.66 	+31 22 45.2 	& SVS 2; LAL 283             	& 0.31 	& -0.81	& 29    & 4		\\
\hline 
\end{tabular}
\end{center}
\end{table*}

As an additional step, we also checked the masked map areas which have low signal-to-noise ratio and inhomogeneous
depth and are therefore not considered useful for accurate photometry. These areas cover most of the remaining
Class II objects catalogued by \citet{2008ApJ...674..336G}. The map contains very few structures, but one 
faint point-source signal is detected, at 14\,mJy, in a region with an RMS of 9\,mJy.

\section{Results}
\label{results}

Disc candidates were found down to 20 mJy, or 2.5 times the per-beam noise. However, 
the distribution of background signals subtracted in the annuli has a standard 
deviation of twice this noise, so weak cold-disc candidates on complex backgrounds need 
to be regarded with caution. Therefore, the aperture photometry could be misleading if there 
is structured emission in the annulus. To estimate the worst-case probability of a false positive
we consider a 1-sigma noise peak lying on a positive background fluctuation at one standard 
deviation, which would produce a nominal 3-sigma candidate (before subtracting the 
annulus signal). This probability is the product of two assumed-Gaussian tails, or P = 
0.159$^2$ = 0.025. For a sample of 34 target positions, it is thus likely that one of 
the disc candidates (from 0.025 $\times$ 34 = 0.86) is not real. Visual checks were 
made to see if each candidate is a plausible compact source. The map with the detected 
and non-detected sources is shown in Figure \ref{map}; the detections are listed in 
Table \ref{t1}, along with their measured 850$\,\mu m$ fluxes and disc mass estimates. 

The flux to disc mass calculation was performed using the formula \citep{2000A&A...358..242S}

\begin{equation}
  M_\mathrm{{disc}} = 0.0188 (1200/\nu)^{3+\beta} F_{\nu} D^2 (e^{0.048\nu/T_\mathrm{{dust}}} - 1)
\end{equation} 

where distance $D$ is in kpc (here adopted as 0.25) and the 
mass is in solar units. It is implicit that the gas-to-dust mass-ratio is 100 and that 
the dust opacity at 1200\,GHz is 0.1 cm$^2$/g, thus making these mass estimates 
compatible with those of cloud cores in NGC~1333 \citep{2001ApJ...546L..49S}
as well as disc studies in other regions \citep[e.g.][]{2009ApJ...694L..36M,2005ApJ...631.1134A}.  
Also from this work, we adopt a dust spectral-index $\beta$ of 1 (typical of the cores 
containing young stars), and a dust temperature of 30 K (within the range 20~K of 
often adopted for cold discs and the 40~K for IRAS-detected objects in NGC 1333). 
These factors yield 0.122 solar masses per Jy at 850$\,\mu m$, or 128 
Jupiter-masses/Jy. The distance estimate is uncertain by $\la 20$ per cent \citep{2013ApJ...775..138S}
and the temperature by around one-third, hence the disc masses are only 
quoted to one significant figure in Table \ref{t1}. More significantly, dust already 
aggregated into `pebble' sizes and above is not detected \citep[e.g.][]{2011MNRAS.412L..88G}.

\begin{figure*}
\includegraphics[width=8cm,angle=-90]{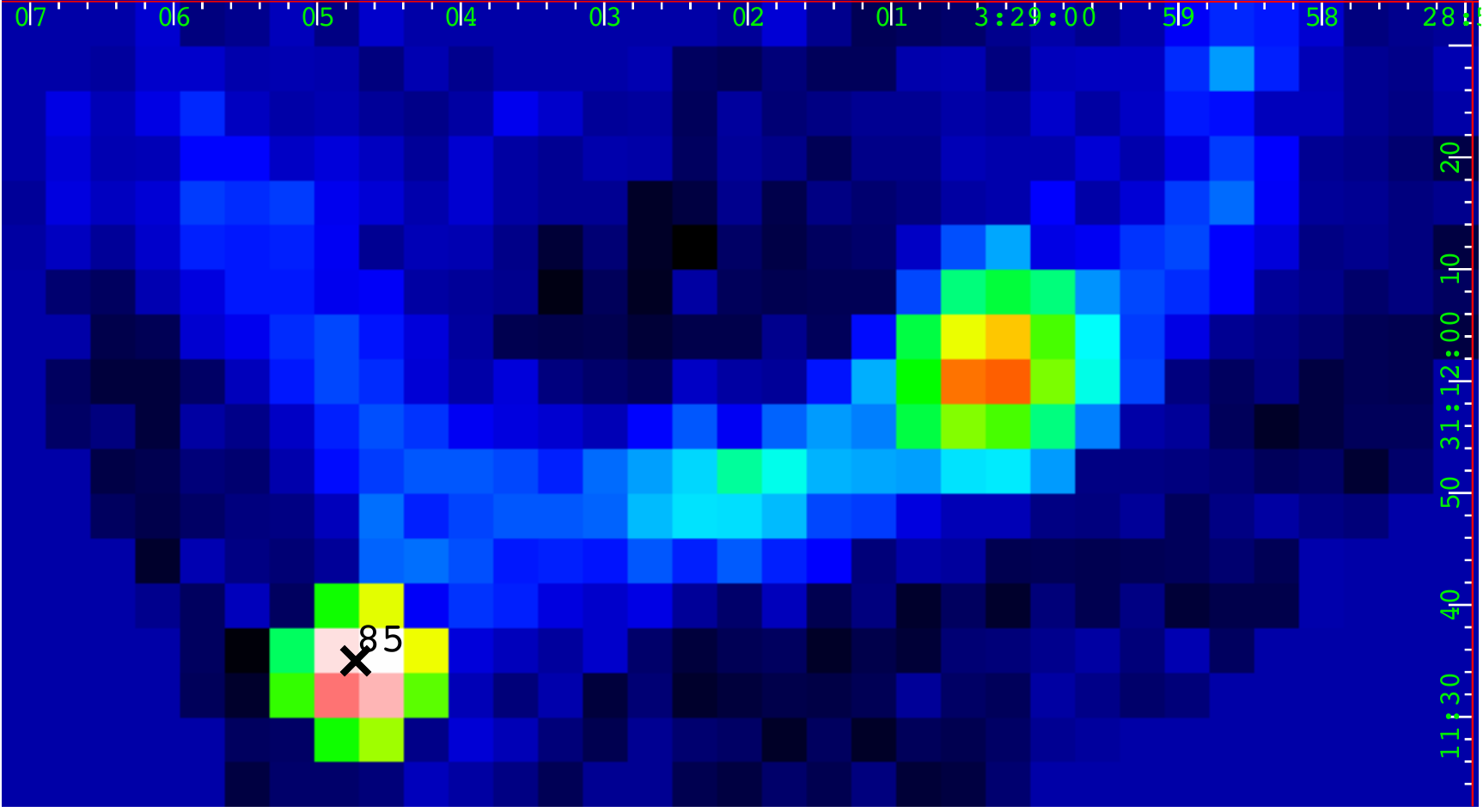}
\caption{850$\,\mu m$ image of a southern region of NGC 1333, with 
a fake source added. The black cross marks the position of the Class II object ASR 99 
(Gutermuth-id\#85) where the fake MMSN-signal was superposed on a real null-signal (see
Sect. \ref{results})
\label{fake}}
\end{figure*}

As pointed out in Sect. \ref{data}, we found one point source detection in the masked areas,
outside the areas considered for reliable photometry. This source has a flux of $\sim 16$\,mJy. 
Given this detection, a MMSN-disc in these areas at 160\,mJy
would have been very prominent. We can therefore rule out the presence of MMSN-discs in this sample
of, in total, 82 Class II objects (94 are catalogued by \citet{2008ApJ...674..336G} 6 are close
to filaments, 6 are too close to the map periphery to be detected).

Thus, the result from this young region is that only one source (1 out of 34) appears to have 
substantial planet-forming potential. None (out of 82) has a disc with sufficient material for the
MMSN. The circumstellar mass-estimate for object \#93 
is 9 Jupiter masses, while all the other disc candidates are 2-4 M$_\mathrm{{Jup}}$, well 
below a 20 Jupiter-mass MMSN. The brightest detection is interesting because the host 
is a candidate sub-stellar object \citep{2009ApJ...702..805S}. However, it has 
relatively high extinction (Table \ref{t1}), so may be at an earlier stage than Class II.
In addition, this source is located in a region heated by B stars which cause higher
dust temperatures \citep{2013MNRAS.429L..10H}. 

We investigated whether MMSN-discs could be present, but missed owing to the high 
dynamic range due to the cloud filaments. Figure \ref{fake} shows flux equivalent to 20 Jupiter 
masses of gas and dust injected into the map after background subtraction, at a position where 
there is a real Class II object but negligible 850$\,\mu m$ emission. The injected flux of 160\,mJy 
was divided over a cross-shape of 12 pixels to simulate spreading over the beam. The four central 
pixels had twice the weight of the eight surrounding ones. Superposing this fake object into 
the real map resulted in a bright detection (Figure \ref{fake}). The injected flux was recovered
to 85\%. This test confirms that a real MMSN could readily be detected even if quite close to a bright 
filament as in Figure \ref{fake}; within the real structure, only 0.3\% of map pixels are in fact 
brighter than this fake peak. The same procedure was repeated, but with injecting the source 
before background subtraction. This yielded similar results (107\% flux recovery).

We note that only one of the sources found in our survey (\#93) is visible in the 450$\,\mu m$ map published
by \citet{2013MNRAS.429L..10H}. This map has a much higher RMS than the map at the longer wavelength
(130\,mJy/beam). With the canonical value of $\beta$ assumed above, all our sources are expected to have
a 450$\,\mu m$ flux below the 2.5\,$\sigma$ limit. For the source \#93 we estimate a relatively high 
submillimeter spectral slope of $\beta = 1.8$, which may confirm that this is a peculiar object, perhaps
in an early evolutionary state.

\section{Discussion}

Our results show that among 34 Class II objects in the mapped 
region of NGC 1333, there is one disc at 9 Jupiter masses and seven 
candidates of 2-4 Jupiter-masses, of which one is likely to be due to 
noise fluctuations as discussed above. Out of a sample of 82 Class II objects,
none has a disc with a mass exceeding the MMSN.
The robustness test of the previous section shows that MMSN discs are unlikely to 
be missed, even in regions of the map with bright residual structure from cloud 
filaments. 

In Table \ref{mmsntab} we compare our results with those of other 
nearby star forming regions, including only Class II sources (i.e. excluding embedded 
Class I as well as diskless Class III objects). Using a variety of literature sources, we 
calculate the fraction of stars per region hosting discs of masses at, or exceeding, 
1\,MMSN, $f_{MMSN}$. These numbers are essentially an updated version of the data 
presented in \citet{2010MNRAS.407.1981G}. In addition,
we derive the fraction of stars in each region hosting discs above our detection limit
of 3\,Jupiter masses, $f_{Mlim}$. For both disc fractions, we calculate consistent 1$\sigma$ binomial
confidence intervals. The table includes regions with ages of 1-2\,Myr (NGC1333,
ONC, Taurus, Ophiuchus, Lupus, Cha-I), 2-5\,Myr (IC348, $\sigma$\,Ori), and 5-10\,Myr
(UpSco), and thus covers the entire age range of Class II sources (see also the revised 
age scale \citet{2013MNRAS.434..806B}). For most of these regions, the disc masses have been 
derived from submm/mm fluxes at 850$\,\mu m$ or 1.3\,mm, using similar assumptions for 
the dust opacity and temperature as well as gas-to-dust ratio; the values should therefore 
be comparable.

For many regions included in Tab. \ref{mmsntab} the submm/mm census is reasonably complete.
This includes Ophiuchus, Taurus-Auriga, IC348 and $\sigma$\,Ori. For other regions, however,
the disc fractions are calculated from an incomplete sample, which raises the issue of potential
biases. In Cha-I the value for the MMSN fraction is based on the more than 20\,yr old study 
by \citet{1993A&A...276..129H}. Their photometry is not homogeneous (the noise varies by a 
factor of 4) and could be biased. However, a more recent assessment of disc masses in Cha-I 
based on Herschel fluxes finds $f_{MMSN}$ of 2-7\% \citep{2014arXiv1405.3833R}, which is in line 
with the previous value. In NGC1333, Lupus and UpSco the sample sizes are also small ($<40$), 
but are without any known bias. In NGC1333 we may miss some discs near the filaments,
as pointed out in Sect. \ref{data}. Strong background emission such as in these filaments is a 
limitation of many submm/mm surveys in star forming regions. For example, the fields observed in 
the ONC \citep{2009ApJ...694L..36M} and IC348 \citep{2011ApJ...736..135L} deliberately avoid
localized regions with cloud emission.

\begin{table*}
\caption{Number of Class II MMSN discs $N$, fraction $f$, and 1$\sigma$ confidence intervals, 
for nearby star forming regions. \label{mmsntab}}
\begin{tabular}{lcccccccl}
\hline
Region        & age (Myr) & $N_{MMSN}$ & $f_{MMSN}$ (\%) & 1$\sigma$(\%) & $N_{Mlim}$ & $f_{Mlim}$ (\%) & 1$\sigma$(\%) & reference \\
\hline 
N1333         & 1-2       & 0/82       & 0  & 0-2   & 8/34	      & 24 & 16-33 & this paper \\
Lupus         & 1-2       & 2/32       & 6  & 2-14  & $\ge$12/32      & 38 & 28-48 & Nuernberger et al. (1997)\\
Cha-I         & 1-2       & 1/14       & 7  & 1-21  & $\ge$6/14       & 43 & 28-59 & Henning et al. (1993) \\
ONC           & 1-2       & 6/55       & 11 & 7-17  & 25/55           & 45 & 38-53 & \citet{2009ApJ...694L..36M} \\
Tau-Aur       & 1-2       & 15/74      & 20 & 16-26 & 38/74           & 51 & 45-58 & \citet{2005ApJ...631.1134A} \\
Oph           & 1-2       & 15/69      & 22 & 17-28 & $\ge$31/69      & 45 & 39-51 & \citet{2007ApJ...671.1800A} \\
IC348         & 2-4       & 0/84       & 0  & 0-2   & 3/84            & 4 & 2-7 & \citet{2011ApJ...736..135L} \\
$\sigma$\,Ori & 3-5       & 0/297      & 0  & 0-1   & $\ge$9/297$^1$  & 3 & 2-4 & \citet{2013MNRAS.435.1671W} \\
UpSco         & 5-10      & 0/37       & 0  & 0-5   & 1/37            & 3 & 0-9 & \citet{2012ApJ...745...23M} \\
\hline
\end{tabular}

$^1$ The detection limit in this study is 4.5$\,M_{Jup}$, i.e. only slightly higher than in our paper.
The quoted lower limit is therefore likely to be close to the actual value.
\end{table*}

\begin{figure*}
\includegraphics[width=15cm]{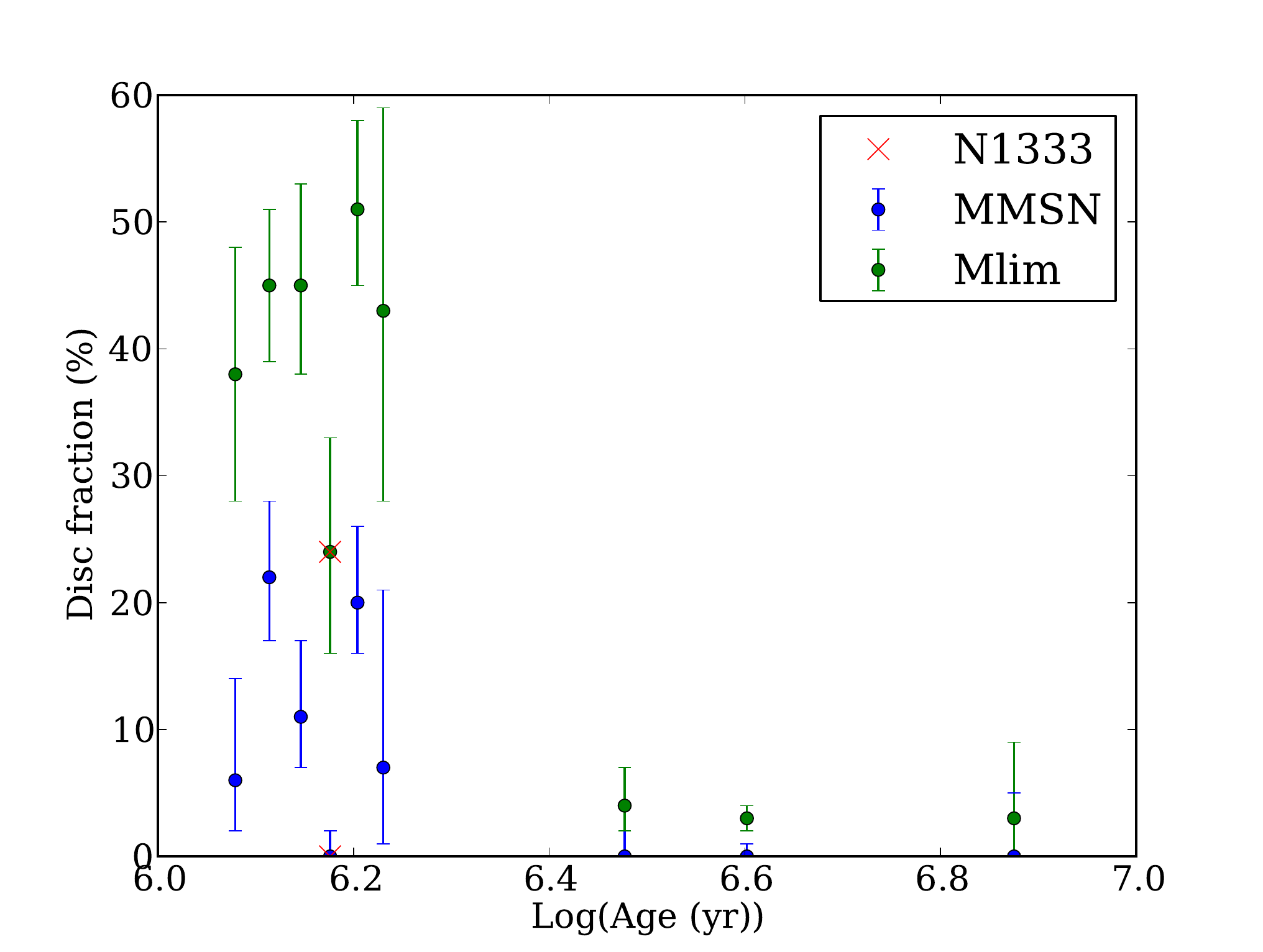}
\caption{Percentages of Class II discs in star forming regions, with masses above MMSN (blue) and
above 3 Jupiter masses (our detection limit, green). Errorbars are binomial 1$\sigma$
confidence intervals. See Table \ref{mmsntab} for references. For the regions with ages between 1-2\,Myr the ages shown
in the plot are arbitrary and have been chosen to allow for a clear presentation. The percentages
for NGC1333 are marked with red crosses. \label{mmsnfig}}
\end{figure*}

The disc fractions in Table \ref{mmsntab} are plotted vs. age of the region in Fig. \ref{mmsnfig}. 
In general only few discs have sufficient dust mass for the formation of massive planetary cores. For 
example, our 2-4 M$_\mathrm{{Jup}}$ candidates in NGC1333 include only about 10 Earth-masses 
of solids, spread over the entire disc. These results are in agreement with the findings from exoplanet
surveys (see Sect. \ref{intro}) which indicate that planetary systems comparable to the solar system
are the exception, and that smaller and less massive systems are more common. Nevertheless, about 20 to 50\% of discs at 
ages of 1-2\,Myr still have sufficient mass to form systems with Super-Earth-type planets.

The figure also demonstrates that $f_{MMSN}$ and $f_{Mlim}$ vary from region to region. IC348, 
UpSco, and $\sigma$\,Ori on the right hand side of the plot contain a 
negligible fraction of MMSN discs and a very low fraction of discs above our mass limit. For these
three regions, the fraction of massive discs is significantly lower than for the 
younger regions.
This has been noted before by \citet{2013MNRAS.435.1671W}, \citet{2011ApJ...736..135L}, as well as 
\citet{2012ApJ...745...23M}. In all three cases the low fraction of massive discs is interpreted 
in the framework of standard viscous evolution of discs; as the objects age, the disc material 
is dissipated and the dust content drops. As a result, the three regions which are slightly 
older than the others are also found to have the lowest fraction of massive discs.

All the remaining regions should have comparable ages, around 1-2\,Myr. In this 
group there is still substantial spread in $f_{MMSN}$ and $f_{Mlim}$. NGC1333, 
the region investigated here, has the lowest value for $f_{MMSN}$ 
and $f_{Mlim}$. The disc fraction in NGC1333 is significantly lower than in
Taurus, Ophiuchus, and the ONC (i.e. with non-overlapping 1$\sigma$ confidence
intervals), and still somewhat lower than in Lupus and Cha-I. Here, age cannot be 
invoked as an explanation. Stars in NGC1333 either already have partially 
formed planetary systems (i.e. a lot of mass is already bound in large grains and
planetesimals) or they will form systems with less massive planets than other regions.
Note that this conclusion is robust against uncertainties in distance or dust temperature; 
even increasing all our disc masses by 50\% would not affect the outcome. 

Thus, this comparison indicates that there are secondary factors, apart from age, that affect the 
fraction of massive discs at Class II stage and thus the potential for planet formation. 
In the following we investigate four possible candidates for such secondary factors.

{\it 1) Dynamical interactions in dense stellar clusters:} 
If this factor were to play a role, we would expect {\it stronger} effects in the ONC, as the 
ONC has higher stellar density than NGC1333 ($\sim 400$\,pc$^{-3}$ vs. $<100$\,pc$^{-3}$,
see \citet{2006ApJ...644..355H} and \citet{2013ApJ...775..138S}). However, stellar 
encounters are unlikely to play a significant role in the disc evolution in the ONC 
\citep[e.g.][]{2001MNRAS.325..449S}, and therefore even less so in NGC1333. Recently, 
\citet{2014MNRAS.441.2094R} use hydrodynamical simulations to show that stellar encounters 
limit disc radii (but not necessarily disc masses) for stellar densities exceeding 
$2-3\times 10^3$\,pc$^{-1}$, more than one order of magnitude higher than in NGC1333. 

{\it 2) Photoevaporation in the UV-radiation field of nearby OB stars:} As in the previous
scenario, we would expect stronger effects in the ONC, which harbours more OB stars than
NGC1333. \citet{2014ApJ...784...82M} find that the ONC is lacking massive discs only in the 
vicinity of the O star $\theta^1$\,Ori C, within 0.03\,pc, whereas in other regions the disc 
mass distribution is similar to regions like Taurus. This is explained by the effect of the 
UV radiation on the discs. NGC1333 does not harbour any O stars, but two B stars 
\citep[B5 and B8, see][]{2013MNRAS.429L..10H}. However, our survey extends to distances of 
0.5\,pc from them, and we do not find any massive discs. Hence, this scenario does not 
seem applicable to NGC1333.

{\it 3) Metallicity:} 
The estimates for the MMSN assume a composition as found in the solar system. A star forming 
environment with a chemical composition different from the solar nebula could yield different planetary 
systems. To our knowledge, the metallicity for stars in NGC1333 is not constrained yet; this scenario
needs further investigation.

{\it 4) Grain growth:} It is conceivable that grains grow faster in NGC1333. In fact, several young protostars 
in NGC1333 have been identified with MMSN-mass discs \citep{2011MNRAS.412L..88G}. Thus, at the very 
early stage, massive discs do exist, but fast coagulation of grains may prevent their detection at
Class II stage. At this point, however, it is not clear what could cause anomalously
fast grain growth. To explore this scenario further, the discs in NGC1333 need to be observed at mm
and cm wavelengths to constrain the grain properties.

\section{Summary and Conclusion}

This is the first wide-field study searching for T~Tauri discs in the submm/mm 
within the NGC~1333 star formation region. Aperture photometry was performed on 
background-subtracted maps. Eight disc candidates were identified. One has 
mass of about one half MMSN, with the rest only hosting a few Jupiter masses of material. 
The fraction of discs with MMSN mass or more in NGC1333 is thus negligible, while the
fraction of discs with masses above our limit of $\sim 3$ Jupiter masses is 24\%.

We compare these values to disc fractions for other star forming regions and find that
an anomalously low fraction of Class II objects in NGC1333 has discs with masses exceeding
the MMSN threshold and discs with masses exceeding our detection limit. We rule out 
age, dynamical interactions and photoevaporation by nearby OB stars as possible reasons
for the lack of massive discs in NGC1333. Other options to explain this result include 
an anomalous metallicity or faster grain growth. Our study thus raises the possibility
that environmental factors (beyond age) can have a significant impact on the evolution
of the discs and the outcome of planet forming processes.

\appendix
\section{The JCMT Gould Belt Survey team}
\label{sect:members}

The current members of the JCMT Gould Belt Survey team are: P. Bastien, S.F. Beaulieu, 
D.S. Berry, H. Broekhoven-Fiene, J. Buckle, H. Butner, M. Chen, H. Christie, A. Chrysostomou, 
A. Chrysostomou, S. Coude, M.J. Currie, C.J. Davis, J. Di Francesco, E. Drabek-Maunder, 
A. Duarte-Cabral, M. Fich, J. Fiege, P. Friberg, R. Friesen, G.A. Fuller, S. Graves, 
J. Greaves, J. Gregson, J. Hatchell, M.R. Hogerheijde, W. Holland, T. Jenness, D. Johnstone, 
G. Joncas, H. Kirk, J.M. Kirk, L.B.G. Knee, S. Mairs, K. Marsh, B.C. Matthews, G. Moriarty-Schieven, 
J.C. Mottram, K. Pattle, J. Rawlings, J. Richer, D. Robertson, E. Rosolowsky, D. Rumble, 
S. Sadavoy, C. Salji, H. Thomas, M. Thompson, N. Tothill, S. Viti, D. Ward-Thompson, 
G.J. White, C.D. Wilson, J. Wouterloot, J. Yates, and M. Zhu.

\end{document}